\documentclass[twocolumns,usenatbib]{mn2e}

\usepackage{amsfonts}
\usepackage{amsmath}
\usepackage{amssymb}
\usepackage{aas_macros}
\usepackage{graphicx}
\usepackage{color}
\usepackage{float}
\usepackage{hyperref}
\usepackage{cleveref}

\def\lsim{~\rlap{$<$}{\lower 1.0ex\hbox{$\sim$}}}
\def\bsim{~\rlap{$>$}{\lower 1.0ex\hbox{$\sim$}}}

\def\apjl{Astrophys.\ J.\ Lett.}

\def\aap{Astron.\ Astrophys.}

\def\apjs{Astrophys.\ J. Supp.}

\def\la{\langle}
\def\ra{\rangle}

\def\ln{{\rm ln}}

\def\mathbi#1{\textbf{\em #1}}

\def\kvh{\mathrm{\hat{\bf{k}}}}
\def\nvh{\mathrm{\hat{\bf{n}}}}

\def\vk{\mathbi{k}}
\def\vp{\mathbi{p}}
\def\vq{\mathbi{q}}
\def\vr{\mathbi{r}}

\def\vv{\mathbi{v}}

\def\vx{\mathbi{x}}
\def\vy{\mathbi{y}}

\newcommand{\abs}[1]{\left\vert#1\right\vert}
\newcommand{\ket}[1]{\left|#1\right\rangle}
\newcommand{\bra}[1]{\left\langle#1\right|}
\newcommand{\Tr}{\textrm{Tr}}

\definecolor{red}{cmyk}{0,1,1,0.55}
\definecolor{blue}{rgb}{0.15, 0.2, .85}

\usepackage[normalem]{ulem}

\begin{document}

\title[Statistics of a single sky]{Statistics of a single sky: constrained random fields and the imprint of Bardeen potentials on galaxy clustering}

\author[Vincent Desjacques, Yonadav B. Ginat and Robert Reischke]
{\parbox[t]{\textwidth}{Vincent Desjacques$^1$, Yonadav Barry Ginat$^1$ and Robert Reischke$^{1,2}$}\\\,\\
$^1$ Physics department, Technion, Haifa 3200003, Israel \\
$^2$ Department of Natural Sciences, The Open University of Israel, 1 University Road, P.O. Box 808, Ra'anana 4353701, Israel}

\date{\today}

\maketitle

\begin{abstract}
  We explore the implications of a single observer's viewpoint on measurements of galaxy clustering statistics. We focus on the Bardeen potentials, which imprint characteristic scale-dependent signatures in the observed galaxy power spectrum. The existence of an observer breaks homogeneity as it singles out particular field values at her/his position, like a spontaneous symmetry breaking. As a result, spatial averaging of the data must be performed while holding the Bardeen potentials fixed at the observer's position. In practice, this can be implemented with the formalism of constrained random fields. In the traditional Cartesian Fourier decomposition, this constraint imprints a signature in the observed galaxy power spectrum at wavenumbers comparable to the fundamental mode of the survey. This effect, which is well within the cosmic variance, is the same for all observers regardless of their local environment because differences of potential solely are observable. In a spherical Bessel Fourier decomposition, this constraint affects the monopole of the observed galaxy distribution solely, like in CMB data. As a corollary, the scale-dependence of the non-Gaussian bias induced by a local primordial non-Gaussianity is not significantly affected by the observer's viewpoint. 
\end{abstract}

\begin{keywords}
cosmology: theory, large-scale structure of Universe, inflation
\end{keywords}

\section{Introduction}

Cosmological inference takes into account the cosmic variance arising from sampling a finite number of modes. It should also take into account the observer's viewpoint or local environment, which induces systematic distortions in statistics of the large scale structure (LSS) relative to an ensemble-averaged measurement made by random cosmological observers.
This effect depends on the correlation length of the physical observable. While galaxy peculiar velocities are noticeably altered \citep{tormen/etal:1993,borgani/etal:2000,hellwing/etal:2017,hellwing/etal:2018}, conditioning on the local density (averaged in a sphere) typically affects density power spectra by less than a percent at cosmological distances, below cosmic variance \citep{reischke/schaefer/etal:2019,hall:2020}.

Recently, \cite{grimm/etal:2020} have studied (among others) the imprint of the Bardeen potentials -- which arise through general relativistic projections \citep[e.g.][]{yoo/etal:2009} and, possibly, a local primordial non-Gaussianity \citep[][]{dalal/etal:2008} -- on the observed galaxy power spectrum $P_g(k)$. Taking into account the dependence of $P_g(k)$ on quantities defined both in the source and the observer rest frame, \cite{grimm/etal:2020} find the latter have dramatic implications as they lead to the cancellation of all ``divergent'' contributions of the form $k^{-4}P_\delta(k)$ and $k^{-2}P_\delta(k)$ induced by the Bardeen potentials (and a few other relativistic terms).
This is quite worrying since a measurement of $f_\text{NL}$ precisely through similar scale-dependent contributions (albeit imprinted by the primordial scalar perturbations rather than the late time Bardeen potentials) has already been performed \citep[e.g.,][]{slosar/etal:2008,giannantonio/etal:2014,leistedt/etal:2014,karagiannis/etal:2014,ho/etal:2015,castorina/etal:2019}, and is an important science goal of forthcoming redshift surveys \citep[e.g.,][]{spherex}.

In this paper, we investigate the implications of a single observer's viewpoint on the existence of ``divergent" contributions in the observed, linear theory galaxy power spectrum. For this purpose, we use the formalism of constrained (conditional) random fields \citep{hoffman/ribak:1991,weygaert/bertschinger:1996}
\citep[see also][for a different approach]{mitsou/etal:2020}. This technique has also been implemented to produce constrained simulations that closely mimic our local environment, including the major nearby structures out to $\sim 100$ Mpc \cite[e.g.,][]{klypin/etal:2003}. We shall focus on the Bardeen potentials owing to their close relation to the non-Gaussian bias, but our considerations apply to any physical quantity arising in $P_g(k)$.

The paper is organized as follows. In \S\ref{sec:constraint}, we briefly review well-known results on constrained random fields, and specialize them to the fluctuation fields contributing to the observed galaxy clustering. In \S\ref{sec:fourier}, we discuss the impact of the observer's viewpoint - or constraint - on their power spectra. In \S\ref{sec:observations}, we apply our findings to the signature of the Bardeen potentials in the observed galaxy power spectrum. We conclude in \S\ref{sec:conclusion}. For convenience, details of the calculation are summarized in an Appendix \S\ref{sec:statistics}.

\section{Constrained random fields}
\label{sec:constraint}

Let $(\eta,\vx)$ be the conformal time and comoving coordinates, and $f(\vx)$ be some homogeneous and isotropic (scalar) random field such as a metric perturbation, the matter density etc. which can be probed with observables like the galaxy distribution or the CMB temperature. For simplicity, we shall assume that $f(\vx)$ is time-independent, and normally distributed with zero mean. We will relax the assumption of Gaussianity in \S\ref{sec:PNG}. 

Consider now an observer $O$ located at the origin of the coordinate system. She/he observes a particular realization of $f(\vx)$ constrained such that $f({\bf 0})=f_o$ (the value of $f_o$ may or may not be locally measurable depending on the nature of $f$). Therefore, if the observable explicitly depends on $f_o$ (an assumption we make throughout this paper), it is essential to regard $f(\vx)$ as a constrained random field (see Appendix \ref{appendix:quantum} for further justification of using a constrained ensemble, rather than an unconstrained one). Since a particular position is singled out, statistics of the constrained field are not homogeneous, but they are still isotropic.

\subsection{Constrained mean and residual field}

As shown in \cite{hoffman/ribak:1991}, $f(\vx)$ can be expressed as the sum of a mean field $\langle f\rangle (r)$, where $r=|\vr|$ denotes the line-of-sight comoving distance, and a residual $\delta f(\vx)$. 
Here and henceforth, all the ensemble averages are formally conditioned to the value $f({\bf 0})=f_o$ of $f(\vx)$ at the origin of coordinate, i.e.
$\langle X \rangle \equiv \langle X|f_o\rangle $.

The mean field is given by
\begin{equation}
\label{eq:meandeltaf}    
  \langle f\rangle (r) = \left(\frac{f_o}{\sigma_f^2}\right) \xi_f(r) \;,
\end{equation}
and is equal to $f_o$ at the origin.
Hereafter, $\sigma_f^2=\langle f^2(\vx)\rangle$ and $\xi_f(r)$ will designate the variance and 2-point correlation function of the {\it unconstrained}, homogeneous and isotropic $f(\vx)$. 

The residual field $\delta f(\vx)$ has zero mean and does not change if $f(\vx)$ is shifted by a constant. Its second-order moment is non-vanishing and independent of the constrained value $f({\bf 0})=f_o$.
More precisely, the variance of the residual depends on the distance $r$ to the observer due to the constraint at the origin, 
\begin{equation}
  \label{eq:vardeltaf}
  \langle \delta f^2\rangle(r) = \sigma_f^2 - \left(\frac{\xi_f(r)}{\sigma_f}\right)^2
\end{equation}
and, in particular, $\langle \delta f^2\rangle(0) = 0$.
This shows that $\delta f(\vx)$ is inhomogeneous, albeit still isotropic. 
However, $\langle\delta f^2\rangle\to\sigma_f^2$ in the limit $r\to\infty$, in accordance with the Cosmological Principle \citep[e.g.][]{milne:1935}. 

The 2-point correlation function $\xi_{\delta f}(\vx_1,\vx_2)$ is the constrained ensemble average $\langle \delta f(\vx_1)\delta f(\vx_2)\rangle$. Since $\delta f(\vx)$ is an inhomogeneous random field, $\xi_{\delta f}=\xi_{\delta f}(\vx_1,\vx_2)$ explicitly depends on both $\vx_1$ and $\vx_2$. For Gaussian statistics, it can be easily evaluated (see \S\ref{sec:statistics}). We find:
\begin{equation}
  \label{eq:xideltaf}
  \xi_{\delta f}(\vx_1,\vx_2) = \xi_f(|\vx_1-\vx_2|) - \frac{1}{\sigma_f^2}\xi_f(r_1)\xi_f(r_2)\;,
\end{equation}
which reduces to \cref{eq:vardeltaf} when $\vx_1=\vx_2$. Moreover, $\xi_{\delta f}\to 0$ whenever any of its argument $\vx_1$ or $\vx_2$ vanishes. 

\subsection{Observed fluctuations}
\label{sec:observedF}

As we shall see in \S\ref{sec:observations}, the fluctuation fields $F(\vx)$ that pertain to the observed galaxy overdensity $\delta_g(z,\nvh)$ are defined by subtracting a mean estimated from the survey itself and, thus, slightly differ from the fields $f(\vx)$ introduced above.
Although there is no unique way of estimating a mean from empirical data, we shall distinguish between two different regimes depending on the characteristics of the survey. In both cases, the ``observed" fluctuation fields $F(\vx)$~\footnote{Only gauge-independent quantities are, strictly speaking, observable} are inhomogeneous, constrained random fields owing to the presence of the observer -- just like $f(\vx)$.

For surveys covering a small area of the sky at effectively one redshift $\bar z$, the mean is given by the observed galaxy number density averaged over the survey volume. In this case, the fluctuation field $F(\vx)$ is generically given by
\begin{align}
    F(\vx) &= {\cal W}(\vx)f(\vx) - \bar f \qquad \mbox{(small survey)} \nonumber \\
    \bar{f} &= \int\!\!\mathrm{d}^3x\, {\cal W}(\mathbf{x})f(\mathbf{x}) \;.
    \label{eq:Fsmall}
\end{align}
Here, ${\cal W}(\vx)$ is the survey window (or mask) function which, for simplicity, we take to be unity for $\vx$ inside the surveyed volume and zero otherwise. This first choice implies that $F(\vx)$ is constrained owing to its dependence on $f(\vx)$.

In the opposite limit of an all-sky survey covering a wide range of redshift, the mean is given by the observed galaxy number density averaged as a function of the observed redshift. In this case, $F(\vx)$ reads
\begin{align}
    F(\vx) &= f(r\nvh) - \bar f_{_\nvh}(r) \qquad \mbox{(all-sky survey)} \nonumber \\
    \bar f_{_\nvh}(r) &= \frac{1}{4\pi}\int\!d\nvh\, f(r\nvh)  \;.
    \label{eq:Flarge}
\end{align}
This second choice implies that the observed fluctuation field $F(\vx)$ vanishes at the origin, like $\delta f(\vx)$. Furthermore, $\bar f_{_\nvh}(r) \to \langle f\rangle(r)$ in the limit $r\to\infty$. Therefore, $F(\vx)$ converges towards $\delta f(\vx)$ in the limit of large separations. At shorter distances however, $F(\vx)\ne \delta f(\vx)$ because $\bar f_{_\nvh}(r)$ will generally differ from $\langle f\rangle(r)$, since only a single realization of the sky is ever observed.

\section{Fourier decomposition}
\label{sec:fourier}

Let us now explore the impact of the constraint in a measurement of the power spectrum of $F(\vx)$. We shall consider a Cartesian Fourier transform approach appropriate to a small survey volume (i.e. small relative to the observable Universe), and a spherical Fourier Bessel decomposition appropriate to an all-sky survey.

\subsection{Cartesian Fourier transform: small survey}
\label{sec:cartesian}

Let $F(\vk)$ be the Fourier mode of the observed fluctuation field $F(\vx)$ defined in \cref{eq:Fsmall}. The Cartesian Fourier transform is defined as 
\begin{equation}
    \label{eq:FourierF}   
    F(\vx) = \int_\vk F(\vk)\, e^{\mathrm{i}\vk\cdot\vx} \;,
\end{equation}
with $\int_\vk \equiv \int \frac{\mathrm{d}^3k}{(2\pi)^3}$. Next, we introduce the traditional power spectrum estimator (which, in practice, is further averaged over spherical Fourier shells to produce a band-power power spectrum) 
\begin{equation}
    V\hat P_F(\vk) = |F(\vk)|^2 \;.
\end{equation}
Here, $V$ is the survey volume. Note that both $F(\vk)$ and $\hat P_F(k)$ have dimension of [Length$^3$].

We need to determine what the empirical power spectrum $\hat{P}_F(k)$ is the unbiased estimator of. Since the Fourier transform of $F(\vx)$ brings about an irrelevant contribution $(2\pi)^3 \delta^D(\vk) \bar f$, where $\delta^D$ denotes a Dirac distribution, $F(\vk)$ essentially is the Cartesian Fourier transform of ${\cal W}(\vx) f(\vx)$. Therefore, $F(\vx)$ is still a constrained random field.
Starting from
\begin{equation}
    \label{eq:Pestimator}
    \langle \hat P_F(\vk)\rangle = \frac{1}{V}\big\langle\big\lvert F(\vk)\big\lvert^2\big\rangle
\end{equation}
where, again, the ensemble average is conditioned to the constraint $f({\bf 0})=f_o$ at the origin, we obtain after some manipulations (the details of which can be found in appendix \ref{sec:statistics}) the desired result
\begin{align}
\langle \hat P_F(\vk)\rangle
    &= \frac{1}{V}\int_\vq P_f(q) \big\lvert{\cal W}(\vk-\vq)\big\lvert^2 + \frac{1}{V\sigma_f^2}\left(\frac{f_o^2}{\sigma_f^2}-1\right) \label{eq:Pestimated} \\
    &\quad \times \int_\vq P_f(q){\cal W}(\vk-\vq)\int_{\vq'}P_f(q'){\cal W}^*(\vk+\vq') 
    \nonumber \;,
\end{align}
where ${\cal W}(\vk)$ is the Fourier transform of the survey window function. This result differs from the unconstrained case  \citep[see, e.g.,][]{feldman/etal:1994,smith/marian:2015} owing to the second, $f_o$-dependent term in the right-hand side.

Equation \eqref{eq:Pestimated} shows that $\hat P_F(\vk)$ estimates the diagonal part of the (pseudo) power spectrum 
\begin{equation}
    \label{eq:Plargevolume}
    \frac{1}{V}P_F(\vk,\vk')=
    \frac{1}{V}\langle f\rangle (k)\langle f\rangle(k') 
    + \frac{1}{V}P_{\delta f}(\vk,\vk') \;,
\end{equation}
where
\begin{align}
\langle f\rangle (k) &= 4\pi \int_0^\infty \!dr\,r^2 \langle f\rangle(r) j_0(kr) \nonumber \\
&= \left(\frac{f_o}{\sigma_f^2}\right)P_f(k) 
\end{align}
is the 1-point, mean field contribution, while 
\begin{equation}
\label{eq:PSdeltaf}    
    P_{\delta f}(\vk,\vk') = (2\pi)^3\delta^D(\vk+\vk') P_f(k) - \frac{1}{\sigma_f^2} P_f(k) P_f(k') \;,
\end{equation}
is the Fourier transform of the connected 2-point correlation \cref{eq:xideltaf} of the residual (see appendix \ref{sec:statistics}).
The constraint $f({\bf 0})=f_o$ at the observer's position implies that $F(\vx)$ is a  non-stationary random field. As a result, the classical Wiener-Khinchin theorem does not hold, because $P_F(\vk,-\vk)$ is not a genuine power spectral density. This is exemplified by the non-vanishing (possibly negative) covariance of $F(\vk)$ and $F(\vk')$ for $\vk'\ne -\vk$.

The variance of $\hat P_F(\vk)$ is given by 
\begin{align}
\label{eq:varPestimated}
    \textrm{var}(\hat{P}_F) &= \langle \hat P_F(\vk)\rangle^2 + \frac{1}{V^2}
    \big\langle F(\vk)^2\big\rangle \big\langle F(-\vk)^2\big\rangle \\
    &\quad - \frac{2}{(V\sigma_f^2)^2}\left(\frac{f_o}{\sigma_f}\right)^4 \bigg[\int_\vq P_f(q) {\cal W}(\vk-\vq)\bigg]^2 \nonumber \\
    &\qquad \times \bigg[\int_\vq P_f(q) {\cal W}^*(\vk+\vq)\bigg]^2  \nonumber \;,
\end{align}
where
\begin{align}
    \frac{1}{V}\langle F(\vk)^2\rangle &= 
    \frac{1}{V}\int_\vq P_f(q) {\cal W}(\vk-\vq){\cal W}(\vk+\vq) \\
    &\quad + \frac{1}{V\sigma_f^2}\left(\frac{f_o^2}{\sigma_f^2}-1\right)
    \bigg[\int_\vq P_f(q) {\cal W}(\vk-\vq)\bigg]^2 \nonumber \;.
\end{align}
For a band-power measurement, \cref{eq:varPestimated} must be further divided by the number of modes $N_k$ in the shell centered around $k$. 

At fixed survey volume $V$, since $N_k\sim k^3V$ (for logarithmic band-power estimates), the variance of a band-power measurement vanishes in the limit $k\to \infty$. Furthermore, ${\cal W}(\vk)$ becomes narrowly peaked around $\vk=0$ as $V$ increases (see appendix \ref{sec:statistics}). Therefore, $\hat{P}_F(k)$ converges (non-uniformly) towards the power-spectrum $P_f(k)$ of the unconstrained field in the limit $k\to\infty$ and/or $V\to\infty$. 
The convergence is non-uniform in the sense that small scales (for which we have many modes) typically converge faster than large scales, although the details of the convergence depend on the shape of $P_f(k)$. In any case, ergodicity holds in the usual sense: spatial averaging of the observed, constrained data is equivalent to ensemble averaging of the unconstrained field for sufficiently large volumes.

\subsection{Spherical Fourier Bessel: all-sky survey}
\label{sec:SFB}

For all-sky surveys, the observed fluctuation field $F(\vx)$ can be conveniently expanded in spherical Bessel functions \citep[see for instance][]{yoo/desjacques:2013}. 
Starting from the angular multipole decomposition
\begin{equation}
    F(\vx) = \sum_{\ell m} a_{\ell m}(r)\, Y_{\ell m}(\nvh)\;,
\end{equation}
where $\nvh=\vx/r$.
and using the Rayleigh expansion, we write the multipole functions $a_{\ell m}(r)$ as
\begin{align}
    a_{\ell m}(r) &= 4\pi i^\ell\int_\vk j_\ell(kr) Y_{\ell m}^*(\kvh) F(\vk) \\
    &= \sqrt{\frac{2}{\pi}}\int_0^\infty\!dk\,k\,j_\ell(kr) F_{\ell m}(k) \nonumber \;.
\end{align}
The last line defines the spherical Bessel transform $F_{\ell m}(k)$ of $F(\vx)$ as
\begin{equation}
    \label{eq:sphericalF}
    F_{\ell m}(k) = \sqrt{\frac{2}{\pi}} \int\!d^3r\, k j_\ell(kr) Y_{\ell m}^*(\nvh)F(\vx)
\end{equation}
and, thereby the spherical power spectrum estimator
\begin{align}
  \hat S_\ell(k,k') &\equiv \frac{2kk'}{\pi}\int\!d^3x_1\int\!d^3x_2\, Y_{\ell m}^*(\nvh_1) Y_{\ell m}(\nvh_2) \nonumber \\
  &\qquad \times j_\ell(kr_1)j_\ell(k'r_2) F(\vx_1) F(\vx_2) 
  \label{eq:Sestimator} \;.
\end{align}
$S_\ell(k,k')$ is independent of the multipole $\ell$ only if the observed fluctuation field is also homogeneous, in which case it simplifies to $\hat S_\ell(k,k')\equiv \hat P_F(k)$. 

We shall now calculate $\langle \hat S_\ell(k,k')\rangle$ for the observed fluctuation field $F(\vx)$ given in \cref{eq:Flarge}. Using $\langle f(\vx_1)f(\vx_2)\rangle = \langle f\rangle(r_1)\langle f\rangle(r_2) + \xi_{\delta f}(\vx_1,\vx_2)$ and subtracting the directional average, we obtain
\begin{align}
    \langle F(\vx_1) F(\vx_2)\rangle &= \xi_f(|\vx_1-\vx_2|)     \label{eq:xiF} \\ 
    &\qquad - \frac{1}{(4\pi)^2}\int\!\!d\nvh\int\!\!d\nvh'\,\xi_f(|r\nvh-r'\nvh'|) 
    \nonumber \;.
\end{align}
Any dependence on $f_o$ has now disappeared.
The first term on the right-hand side of \cref{eq:xiF} yields a contribution 
\begin{equation}
\frac{2}{\pi^2}\int_0^\infty\!\!dq\,q^2 P_f(q) \mathcal{M}_\ell(k,q) \mathcal{M}_\ell(k',q)
\end{equation}
to the spherical power spectrum $S_\ell(k,k')$, where
\begin{equation}
\mathcal{M}_\ell(k,q) = k \sqrt{\frac{2}{\pi}}\int_0^\infty\!\!dr\,r^2 j_\ell(kr) j_\ell(qr) \;.
\end{equation}
Similarly, on applying the identity
\begin{equation}
    \int_0^\infty\!\!dr\,r^2 j_0(kr) j_0(qr) = \frac{\pi}{2kq}\,\delta^D\!(k-k')\;,
\end{equation}
the directional average piece leads to the simple expression
\begin{equation}
\label{eq:Smonopole}
    P_f(k)\, \delta^D\!(k-k')\, \delta^K_{\ell 0} \;,
\end{equation}
where $\delta^K$ designates the Kronecker symbol. Therefore, \cref{eq:Sestimator} is an unbiased estimate of
\begin{align}
    S_\ell(k,k') &=\frac{2}{\pi^2}\int_0^\infty\!\!dq\,q^2 P_f(q) \mathcal{M}_\ell(k,q) \mathcal{M}_\ell(k',q) \nonumber \\
    &\quad + P_f(k)\, \delta^D\!(k-k')\, \delta^K_{\ell 0} 
    \label{eq:Sestimated} \;,
\end{align}
The second term in the right-hand side contributes only to the diagonal part of the monopole power spectrum $S_0(k,k')$. The latter cannot be properly estimated from the data since we can observe only one sky and, thus, measure one realization of $F_{00}(k)$ solely. 

Note that, unlike $S_0(k,k')$, any realization of $F_{00}(k)$ will depend on $f_o$, in analogy with the dependence of the CMB monopole on the value of the gravitational potential at the observer's position. Therefore, the knowledge of $P_f(k)$ (and, realistically, the survey window function) can be used to determine $f_o$ from a measurement of $F_{00}(k)$. This is consistent with the fact that $f_o$ can, in principle, also be determined from a Cartesian Fourier decomposition as discussed above.

\section{Galaxy clustering}
\label{sec:observations}

In this Section, we apply our previous results on constrained random fields to the signature of the Bardeen potentials in the observed galaxy distribution. In the conformal Newtonian gauge adopted here, they are the gravitational potential $\psi(\vx)$ and the spatial curvature $\phi(\vx)$. 

\subsection{Metric and observer}

Focusing on scalar perturbations in the linear regime, the perturbed FLRW metric (around a spatially flat background) is given by
\begin{equation}
    ds^2 = a^2(\eta)\big[-(1+2\psi)d\eta^2 + (1-2\phi)dx_i dx^i\big] \;.
\end{equation}
The Fourier modes of the unconstrained potentials $\psi=\psi(\eta,\vx)$ and $\phi=\phi(\eta,\vx)$ can be expressed in terms of the comoving curvature perturbation $\zeta(\vx)$. For instance, we have  $\psi(\eta,\vk)=(3/5)T_\psi(\eta,k)\zeta(\vk)$ with a transfer function given by $T_\psi(\eta,k)=g(\eta)T(k)$.
Therefore, $\psi$ and $\phi$ are generally not homogeneous functions of time. For simplicity however, we will ignore the time-dependent growth factor $g(\eta)$ (which is constant in an EdS universe). As a result, the unconstrained potentials are statistically homogeneous and isotropic at all time $\eta>\eta_*$, where $\eta_*$ is the epoch of last-scattering. Taking into account $g(\eta)\ne 1$ at late time would not change any of our conclusions. 

Furthermore, we consider an observer $O$ located at the origin of the coordinate system, and moving with a 4-velocity $u=a^{-1}(1-\psi_o,\vv_o)$. Owing to the presence of $O$, $\psi(\vx)$ and $\phi(\vx)$ now are constrained random field which assume a particular value $\psi_o$ and $\phi_o$ at the origin (and so does the velocity field $\vv$). 
 
\subsection{Galaxy number counts}
\label{subsec: Galaxy number counts}

In linear theory, the observed fluctuation in galaxy number counts is given by an intrinsic contribution $b_1\delta$ plus a number of projection effects which arise through the redshift between the source and observer, gravitational lensing and, no less importantly, a distortion of the physical volume at the source position. 

Following the computation of \cite{challinor/lewis:2011} (carried out in the conformal Newtonian gauge), the observed galaxy number counts as a function of observed redshift $z$ and position $\nvh$ on the sky is
\begin{align}
\label{eq:GRng}
    n_g(z,\nvh) &= \frac{a^3\,\bar r^2\,\tilde n_g}{{\cal H}(1+z)}\bigg\{1+b_1\delta^\text{\tiny sc} + \frac{\dot{\tilde{n}}_g}{\tilde n_g}\delta\eta+2\frac{\delta r}{r}+2{\cal H}\delta\eta-2\kappa \nonumber \\
    &\qquad +3 \nvh\cdot\vv_o -\left(\frac{\dot{\cal H}}{\cal H}-{\cal H}\right)\delta\eta-\frac{1}{\cal H}\frac{d\psi}{d\eta}+\frac{1}{\cal H}\big(\dot{\psi}+\dot{\phi}\big) \nonumber \\
    &\qquad +\frac{1}{\cal H}\nvh\cdot\frac{d\vv}{d\eta}+\psi-2\phi+\nvh\cdot\big(\vv-\vv_o\big) \bigg\}
\end{align}
in linear theory.
Here, $\bar r=\bar r(z)$ is the line-of-sight comoving distance in the unperturbed FRW background, $b_1$ is the linear galaxy bias \citep{kaiser:1984}, and $\tilde n_g=\tilde n_g(z)$ is the average, rest-frame galaxy number density at the observed redshift $z$ (that is, averaged over all the possible observer configurations). 
All the fields on the right-hand side are evaluated at spacetime position $(\bar\eta(z),\bar r(z)\nvh)$, where $\bar\eta(z)$ is the time coordinate in the background and $\nvh$ is the observed direction on the sky. Furthermore, the matter overdensity $\delta^\text{\tiny sc}$ is in a synchronous gauge (comoving with matter), so that $\delta$ and $\psi$ are related through the usual Poisson equation.
Finally, $\delta\eta$ and $\delta r$ are the time and radial coordinate fluctuations on hypersurfaces of constant observed redshift.

Note that, while the redshift fluctuation involves differences of quantities evaluated at the source and observer's position such as $\psi-\psi_o$ (which vanish at the origin), the volume distortion brings about terms that depend only on $\phi$ and $\psi$ at the source position (e.g. $\psi-2\phi$ in the third line of \cref{eq:GRng}. 
Therefore, defining the observed galaxy fluctuation field $\delta_g(\nvh,z)$ relative to the ``global mean'' $\frac{a^3\bar r\tilde n_g}{{\cal H}(1+z)}$ as done in \cite{challinor/lewis:2011} implies that the volume distortion can be locally measured if the mean galaxy density is known. Alternatively, the observer $O$ can define $\delta_g(\nvh,z)$ relative to the observed mean counts and locally measurable quantities, as emphasized in \S\ref{sec:observedF}. Proceeding along these lines, one eventually obtains \citep[see][for a detailed derivation and discussion]{ginat/etal:2021}
\begin{align}
    \label{eq:deltagobs}
    \delta_g(\nvh,z) &= b_1\delta^\text{\tiny sc}+\frac{\dot{\tilde{n}}_g}{\tilde n_g}\delta\eta
    +\left(3{\cal H}-\frac{\dot{\cal H}}{\cal H}\right)\delta\eta+2 \frac{\delta r}{r}
    + \psi - 2\phi \nonumber \\
    &\quad -3 \left(\psi_o+\frac{1}{{\cal H}_o}\dot{\phi}_o\right) + \frac{1}{{\cal H}_o}\partial_i v_o^i
    -\frac{1}{\cal H}\frac{d\psi}{d\eta}+\frac{1}{\cal H}\big(\dot{\psi}+\dot{\phi}\big)
    \nonumber \\
    &\quad -2 \big(-\nvh\cdot\vv_o+\kappa\big)+\frac{1}{\cal H}\nvh\cdot\frac{d\vv}{d\eta} + \nvh\cdot\vv_o \;,
\end{align}
which ensures that $\delta_g(\nvh,z)$ only involves differences of potential, i.e. $\psi-\psi_o$ and $\phi-\phi_o$.

\subsection{Small survey}

For a small spectroscopic survey, $O$ will estimate the mean galaxy number density as
\begin{equation}
    \bar n_g = \int\!\!\mathrm{d}z\int\!\!\mathrm{d}\nvh\, {\cal W}(z,\nvh) n_g(z,\nvh) \;.
\end{equation}
By assumption, the survey window function ${\cal W}(z,\nvh)$ is non-vanishing solely in a small interval centred at the effective redshift $\bar z$ and in a small patch of the sky. 
The observed galaxy fluctuation field $\delta_g(z,\nvh)$ thus is
\begin{equation}
    \label{eq:deltagobssmallsurvey} 
    \delta_g(z,\nvh) = \frac{n_g(z,\nvh)}{\langle n_g\rangle}-1
\end{equation}
with $\langle n_g\rangle$ being the observed average galaxy density,
and depends on quantities such as, e.g.,
\begin{equation}
\label{eq:PsiObs}
\Psi(\vx) = {\cal W}(\vx)(\psi(\vx) -
\psi_o) - ({\bar \psi}-\psi_o)
\end{equation}
analogous to \cref{eq:Fsmall}. In a Cartesian Fourier analysis, the estimated galaxy power spectrum $\hat P_g(\vk)$ thus involves the auto- and cross-power spectra of these fluctuation fields. In particular, $\hat P_g(\vk)$ includes a term proportional to $\hat P_\Psi(\vk)$, the ensemble average of which is obtained upon replacing $F\to \Psi$ and $f\to\psi-\psi_o$ in \cref{eq:Pestimated}, so that $f_o= 0$ and, consequently, $\Psi_o=0$.

To get insight into the impact of the observer constraint on $\hat P_\Psi(\vk)$, consider first a sufficiently large volume $V$ such that ${\cal W}(\vk)$ is narrowly peaked at $\vk=0$. We thus have
\begin{equation}
\label{eq:Pestimated_nowindow}
    \langle\hat P_\Psi(\vk)\rangle \approx P_\psi(k) - \frac{1}{V\sigma_\psi^2} P_\psi(k)^2 \;.
\end{equation}
We seek to estimate the magnitude of the second term (in the right-hand side) relative to the first, that is, $\sim (V\sigma_\psi^2)^{-1}P_\psi(k)$. 
For this purpose, we assume a nearly scale-invariant spectrum (of primordial scalar fluctuations) such that the unconstrained power spectrum is $P_\psi(k) \propto k^{-4+n_\mathrm{s}} T^2(k)$, where $n_\mathrm{s} = 0.963$ is the scalar spectral index and $T(k)$ is usual (late time) transfer function normalized to unity in the limit $k\to 0$.
Furthermore, we introduce an infrared cutoff $k_L$ in the computation of $\sigma_\psi^2$ \citep[see also][]{bertacca/eta:2012,raccanelli/etal:2014}. This cutoff reflects the fact that the long mode $\psi_L\equiv \psi(k<k_L)$ is unobservable and, as such, absorbed into the observed background cosmology (which slightly differs from the global FLRW spacetime). As an order of magnitude estimate, we have $k_L\sim {\cal O}(1/r_*)$ where $r_*$ is the line-of-sight comoving distance to the last-scattering surface. Taking into account the suppression of the transfer function $T(k)$ for $k\gtrsim k_\text{eq}$, we find (assuming $|n_s-1|\ll 1$)
\begin{equation}
    \label{eq:PPsiEffect}
    \frac{P_\psi(k)}{V\sigma_\psi^2} \sim \frac{1}{4\pi} \left(\frac{k_V}{k}\right)^3 
    \ln\!\left(\frac{k_\text{eq}}{k_L}\right)^{-1}\left(\frac{V}{V_L}\right)^{\frac{1-n_s}{3}} \;.
\end{equation}
Here, $k_V = 2\pi/V^{1/3}$ is the fundamental mode defined by the survey volume and $V_L$ is the comoving volume of the observable Universe. This demonstrates that, on the survey scale $k_V$, the constraint's effect is proportional to the inverse of ln($\frac{k_\text{eq}}{k_L}$), and it decays like $(k_V/k)^3$ for $k>k_V$. Furthermore, the effect vanishes in the limit $k_L\to 0$.

\begin{figure}
\begin{center}
\includegraphics[width = 0.5\textwidth]{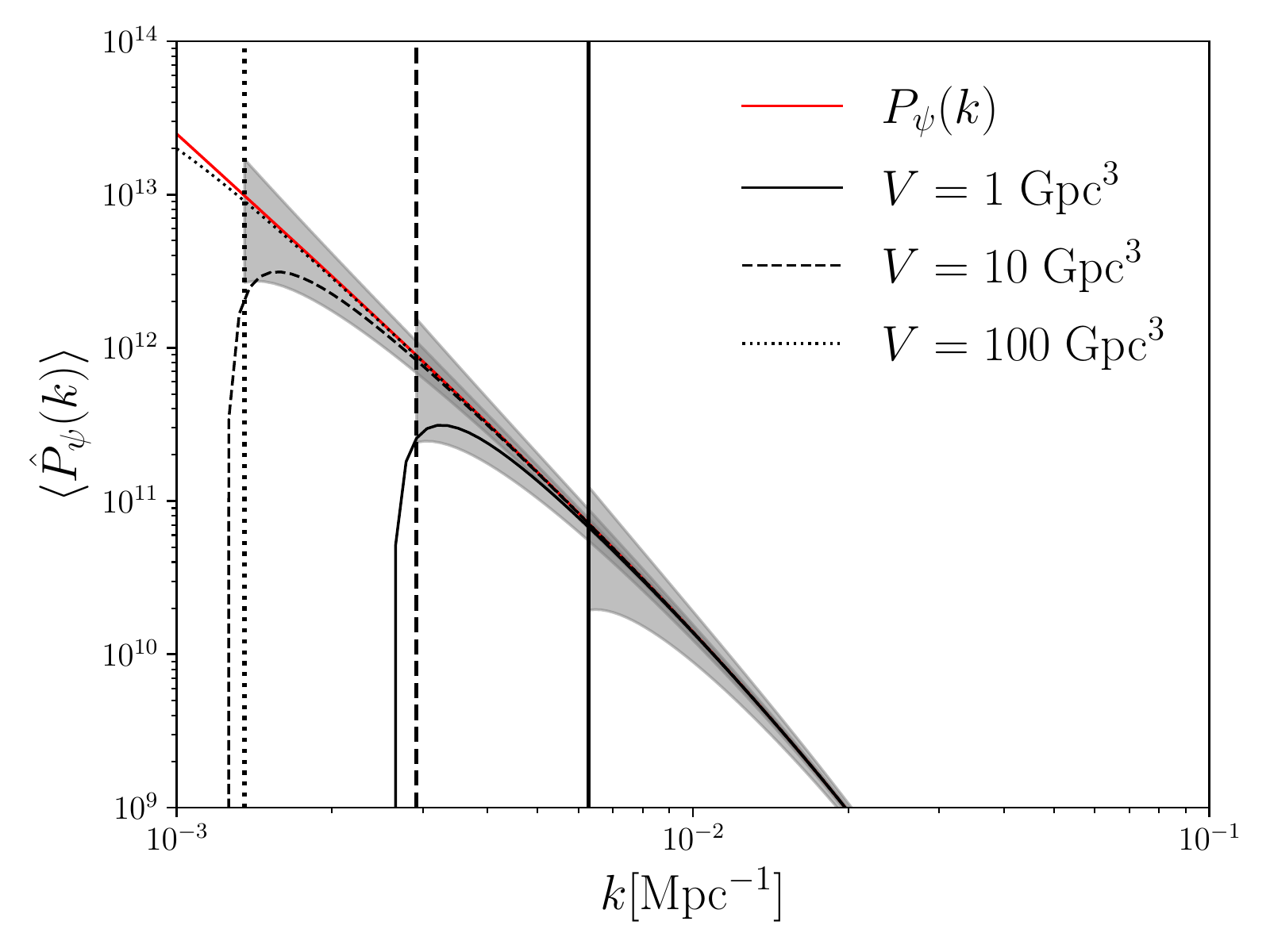}
\caption{Estimator of the power spectrum for different survey sizes, \cref{eq:Pestimated}. The red line corresponds to the unconstrained field. 
The fluctuation field $\Psi$ defined in \cref{eq:PsiObs} is constrained such that $\Psi_o=0$ at the observer position.
Vertical lines indicate the fundamental mode $k_V$ of each survey. Only the wavemodes with $k>k_V$ are measured from a given survey. The shaded region represents the cosmic variance up to the corresponding fundamental mode.}
\label{fig:relative_contribution}
\end{center}
\end{figure}

For the sake of completeness, \cref{fig:relative_contribution} displays the effect of the constraint $\Psi_o=0$ on the ensemble average $\langle\hat P_\Psi(\vk)\rangle$ computed from  \cref{eq:Pestimated} using the spherical mask function given in Appendix \S\ref{sec:statistics}. 
The term $(\frac{\psi^2_o}{\sigma^2_\psi}-1)$ becomes $-1$ when $f$ is replaced by $\psi-\psi_o$.
The resulting, shell-average $\langle\hat P_\Psi(k)\rangle$ is shown for three different survey volumes (solid, dashed and dotted black line), and compared to the unconstrained spectrum $P_\psi(k)$ (solid red line). The vertical lines mark the fundamental mode of each survey. For the infrared cutoff $k_L$, we take a conservative (high) value corresponding to a comoving scale of $10\;\mathrm{Gpc}$. \cref{fig:relative_contribution} shows that the observer's viewpoint leaves an imprint on $\langle\hat P_\Psi(k)\rangle$ only for $k\sim k_V$ where cosmic variance is large. This agrees with our analytic estimate \cref{eq:PPsiEffect}. 
Furthermore, \cref{fig:relative_contribution} exemplifies the non-uniform convergence to the unconstrained statistics: the constraint's imprint increases with decreasing wavenumber.

\subsection{All-sky survey}

For the all-sky survey, $O$ estimates the mean galaxy number density according to
\begin{equation}
\label{eq:observedmean}
    \bar n_g(z) = \frac{1}{4\pi}\int\!d\nvh\,n_g(\nvh,z)
    \qquad \mbox{(all-sky survey)}
\end{equation}
as in \cref{eq:Flarge}. 
As a result, $\bar n_g(z)$ depends on directional-average quantities such as $\bar\psi_{_\nvh}(r)$ etc. where
\begin{equation}
    \bar\psi_{_\nvh}(r) = \frac{1}{4\pi}\int\!d\nvh\,\psi(r\nvh)
\end{equation}
converges towards the mean field $\langle\psi\rangle(r)$ (conditioned to $\psi=\psi_o$ at the origin) in the limit $r\to\infty$ solely.
Consequently, since $\psi-\bar\psi_{_\nvh}(r)=\Psi-\bar\Psi_{_\nvh}(r)$ where $\Psi(r\nvh)=\psi(r\nvh)-\langle\psi\rangle (r)$ is the residual gravitational potential, the observed galaxy fluctuation field 
\begin{equation}
    \label{eq:deltagobsallsky}
    \delta_g(z,\nvh) = \frac{n_g(z,\nvh)}{\bar n_g(z)}-1 
\end{equation}
will take the same form as the fractional perturbation to the global mean in Eq.~(\ref{eq:GRng}), albeit with $\psi$, $\phi$ etc. replaced by 
\begin{equation}
    \psi \to \Psi - \bar \Psi_{_\nvh}(r)\;,\quad 
    \phi \to \Phi - \bar \Phi_{_\nvh}(r)\quad \mbox{etc.}
\end{equation}
The presence of $\bar\Psi_{_\nvh}(r)$ etc. originates from the fact that the directional average moves a piece of the fluctuations into the observed mean $\bar n_g(z)$. 

The spherical Fourier Bessel decomposition presented in \S\ref{sec:observations} can be applied to $\delta_g(z,\nvh)$ to yield the galaxy multipoles $\Delta_{\ell m}^g(k)$, with $\Psi(\vx)$ etc. playing the role of $F(\vx)$. The observed spherical power spectra $\hat S_\ell(k,k')$ will thus involve the power spectra $P_\psi(k)$ etc. of the unconstrained fields. Therefore, the results of \cite{yoo/desjacques:2013} are perfectly valid: the gravitational potential also leads to large-scale, divergent contributions of the form $k^{-4}P_\delta(k)$ etc. in the observed galaxy spherical power spectrum. Moreover, apart from the monopole $\Delta_{00}^g(k)$, all the power spectra $\hat S_\ell(k,k')$ are independent of the value $\psi_o$ etc. at the observer's position.

Note that, in practice, any all-sky survey will involve a non-trivial mask function, from which the measured $\hat S_\ell(k,k')$ must be deconvolved in order to recover the contribution proportional to $k^{-4}P_\delta(k)$ etc. We have ignored this complication because it does not affect any of our conclusions. 

\subsection{Primordial non-Gaussianity}
\label{sec:PNG}

Our considerations also apply to the non-Gaussian bias induced by a primordial non-Gaussianity. For a local bispectrum shape, the observed number counts include a term $f_\text{NL}b_\phi \phi_i(\vx)$ (in linear theory), where $b_\phi$ is the non-Gaussian bias and $\phi_i(\vx)=(3/5)\zeta(\vk)$ is the spatial curvature deep in matter domination \citep{dalal/etal:2008}. Our analysis implies that a local primordial non-Gaussianity also leads to contributions of the form $b_\phi^2k^{-4}P_\delta(k)$ and $b_\phi k^{-2}P_\delta{k}$ (for $k\lesssim k_\text{eq}$) in the observed (Cartesian or spherical) galaxy power spectrum.

A local primordial non-Gaussianity also affects the constrained statistics of the Bardeen potentials which, thus far, have been computed under the assumption of Gaussianity. As we demonstrate in Appendix \S\ref{sec:statistics} however, this effect has an amplitude $\lesssim 10^{-4}$ relative to the Gaussian contribution to the covariance $\langle \phi_i(\vk)\phi_i(\vk')\rangle$ for admissible values of $f_\text{NL}\sim 1$ and, therefore, can be safely neglected.

\section{Discussion}
\label{sec:discussion}

An observer measures $\hat{P}$ (or $\hat{S}_\ell$) averaged over a $k$-band (or the multipoles $-\ell\leq m\leq \ell$). The constraint $f({\bf 0}) = f_o$ induces a correlation between the local environment of the observer and the observed data. As a consequence, the observed wavemodes $F(\vk)$ are not independent, and one must rely on constrained statistics. The correct probability measure thus is that of the constrained random field. This implies that the measured $\hat{P}$ is related to the theoretical $\hat{P}$ computed in the constrained ensemble.

In contrast to our approach, both \cite{mitsou/etal:2020} and \cite{grimm/etal:2020} do not take into account a constraint $f({\bf 0}) = f_o$ and, thus, the fact that the probability ensemble is constrained. Although they consider only the limit $V\to\infty$ in which ergodicity holds -- so that spatial averaging of the observed constrained data is equivalent to ensemble averaging of the unconstrained field as we demonstrated above -- only the fields at the source position become unconstrained in that limit. Therefore, one should never average over $f_o$ regardless the value of $V$ and, in particular, when "$V = \infty$" although, in this limit, the constraint is completely diluted.

Another important difference with \cite{grimm/etal:2020} \citep[and][who considered the luminosity distance]{biern/yoo:2017} is the explicit dependence of our results on an infrared cutoff $k_L$. In our approach, $k_L$ corresponds to the largest fluctuations measurable by the observer, that is, the size of her/his cosmological horizon. In other words, $k_L$ is not arbitrary: it is another physical scale which the effect is expected to depend on. The long mode $\psi_L$ constructed from the wavemodes with $k<k_L$ is absorbed into an effective cosmology \citep[along the lines of][]{baldauf/etal:2011}, so that the constraint really is imposed on the short-wavelength piece of $\psi(\vx)$ (the fluctuating piece) constructed from the Fourier modes with $k>k_L$.
Treating the observable patch as a tophat sphere of radius $\sim 1/k_L$ is, of course, a simplification introduced for convenience only. A realistic calculation would take into account the CMB visibility function etc. Since the dependence on $k_L$ is weak (it is within the cosmic variance for typical fluctuations), we do not expect our conclusions to depend on the details of such an implementation. 
By contrast, \cite{grimm/etal:2020} define $k_L$ (i.e. $k_\text{IR}$ in their notation) such that $k_L\bar r\ll 1$, which implies that $k_L < k_V$. However, the "divergent" part (this term is a bit of a misnomer since both $\psi$ and $\phi$ asymptote to constant on superhorizon scales) of the observed power spectrum arises from wavemodes $k\gtrsim k_V$ close to the survey boundary. Therefore, the argument given by \cite{grimm/etal:2020} does not demonstrate there are no $k^{-4}P_\delta(k)$ etc. contributions as they do not consider the effects of (short) modes close to the survey fundamental mode.

To conclude this discussion, note that the restriction imposed by the observer viewpoint can also be motivated from a quantum mechanical perspective. We refer the interested reader to Appendix \ref{appendix:quantum} for more details.

\section{Conclusion}
\label{sec:conclusion}

We have explored the impact of an observer's viewpoint (or local environment) on the observed galaxy power spectrum using the formalism of constrained random fields \cite[e.g.][]{hoffman/ribak:1991}. We have focused on the Bardeen potentials, which imprint a characteristic signature $\propto k^{-2}\delta(\vk)$ in the galaxy number counts.
Our findings can be summarized as follows: 
\begin{itemize}
\item The existence of an observer effectively breaks the homogeneity of the underlying fluctuation fields, like a spontaneous symmetry breaking. Therefore, it is essential to perform the averaging from the vantage point of a particular observer, rather than averaging over randomly positioned observers. The symmetry is that of placing the observer anywhere in the Universe -- it is spontaneously broken by choosing to place the observer at a single point (which must be the case, as we can only observe the Universe from a single point in space). Any statistical analysis done by the observer must incorporate this fact. 
\item The viewpoint of a single observer can be characterized by assigning specific values at her/his position to cosmological fluctuation fields. This applies also to the Bardeen potentials $\psi(\vx)$ and $\phi(\vx)$. Consequently, spatial averaging of the data must be performed while holding $\psi(\vx)$ and $\phi(\vx)$ fixed to $\psi_o$ and $\phi_o$ at the observer's position. The symmetry -- of placing the observer at a point with any value e.g. $\psi_o$ of $\psi(\vx)$ -- is spontaneously broken by focusing on a single observer in a single Universe. 
\item The observer's viewpoint constrains the observed galaxy power spectra, yet the signature depends on the survey characteristics and the method of analysis. For the traditional Cartesian decomposition of a small survey, the constraint imprints a signature only for wavenumbers close to the fundamental mode. For a spherical Fourier Bessel decomposition of an all-sky survey, the constraint affects the monopole of the observed fluctuation field solely, analogously to the CMB \cite[see, e.g.][for a related discussion in the CMB context]{zibin/scott:2008}. 
\item For the contributions to the observed galaxy power spectrum that arise from Bardeen potentials, an infrared cutoff $k_L$ must be introduced for the evaluation of the constraint in order to account for the unobservability of the long mode. Notwithstanding, the impact of the constraint (in a Cartesian analysis) weakly depends on $k_L$ and, moreover, its magnitude is significantly smaller than cosmic variance. Therefore, the expected $k^{-4}P_\delta(k)$ etc. contributions to the observed galaxy power spectrum are recovered. 
\item Ergodicity holds in the sense that, in the limit of large survey volumes $V$, constrained spatial averaging is equivalent to ensemble averaging of the unconstrained field. However, the constrained ensemble is different from the unconstrained one for any finite value of $V$.
\item Only the difference $\psi-\psi_o$ ($\phi-\phi_o$) of gravitational potential (spatial curvature) can be extracted from galaxy clustering data and, generally, from any large scale structure data, in agreement with the intuition that adding a constant to the potential (in the weak field limit) has no observable consequences \citep{ginat/etal:2021}. This also applies to the modulation induced by super-survey gravitational potential fluctuations \cite[see][]{castorina/dizgah:2020,darwish/etal:2020}.
\end{itemize}
Therefore, we reach conclusions different from \cite{grimm/etal:2020} mainly because we do not average over the contributions at the observer's position which, we contend, is unjustified. They are fixed (and do not diverge) when spatial averaging of the data is performed and, thus, must be implemented as a constraint. The observer's viewpoint constrains any fluctuation field to assume either some fixed, locally measurable value $f_o\ne 0$ (as is the case of, e.g., the peculiar velocity field $\vv$ relative to the CMB frame), or $f_o\equiv 0$ if the perturbation cannot be measured locally (as is the case of, e.g., the Bardeen potentials $\psi$ and $\phi$). However, $\psi-\psi_o$ etc. could be determined locally with future observations of structures in the nearby Universe. Constrained simulations of the local LSS should be useful for this purpose.

Finally, as a corollary of our findings, the scale-dependence of the non-Gaussian bias is not affected by the observer's viewpoint and, thus, previous results in the literature remain perfectly valid.

\vspace{2mm}

{\bf Data Availability}: The data underlying this article will be shared on reasonable request to the corresponding author.

\section{Acknowledgments}

We thank an anonymous referee for a helpful and constructive report.
We acknowledge support by the Israel Science Foundation (grant no. 1395/16).  R.R acknowledges additional support by the Israel Science Foundation (grant no. 255/18).

\bibliographystyle{mn2e}
\bibliography{references}

\appendix

\section{Constrained statistics}
\label{sec:statistics}

We use a path integral approach to calculate statistics of constrained random fields. We refer the reader to \cite{hoffman/ribak:1991,binney/quinn:1992,weygaert/bertschinger:1996} for other treatments. The advantage of the formalism we employ here is that it can easily incorporate non-Gaussianity, and can be used to compute all $n$-point functions.  

\subsection{Power spectrum}
\label{subsec:power-spectrum constrained}

Let $f(\vx)$ be an underlying homogeneous and isotropic field.
We wish to compute the constrained power spectrum $\langle f^*(\vp) f(\vq)\rangle$, 
where the ensemble averages should again be understood as $\langle X \rangle = \langle X |f_o\rangle$.  

Since $\langle f\rangle$ and $\delta f(\vx)$ are statistically uncorrelated, $\langle f^*(\vp) f(\vq)\rangle$ decomposes into the sum of $\la f\ra (p) \la f\ra (q)$ and the covariance $\la \delta f^*(\vp) \delta f(\vq)\ra$ of the residual.
Focusing on the latter (the former is trivial), we can set $f_o=0$ without any restriction since the residual field $\delta f(\vx)$ does not depend on the value of $f_o$ \citep[see][]{hoffman/ribak:1991}.

Up to an overall normalisation, the power spectrum conditioned to $f_o=0$ is given by the path integral
\begin{equation}
    \label{eq:pathintegral}
  \int \mathcal{D}f\, \delta^D\!\!\left(\int_\vk f(\vk)\right)f^*(\vp)f(\vq)\,e^{-\int_\vk\!
  \frac{\abs{f(\vk)}^2}{2P_f(k)}} \;.
\end{equation}
This path integral arises from the infinite-dimensional generalisation of the finite-dimensional partition function (for a finite volume, with a finite number of independent Fourier modes)
\begin{equation}
    Z\sim\left[\prod_{\vr}\int \mathrm{d}f_\vr\right] \, \delta^D\!\!\left(\sum_{\vk} f_\vk\right)\,e^{-\sum_\vk\!
  \frac{\abs{f_\vk}^2}{2P_f(k)}} \;,
\end{equation}
where $\sum_\vk\!\frac{\abs{f_\vk}^2}{2P_f(k)}$ can be thought of as a free energy. The Dirac functional $\delta^D\!\!\left(\sum_{\vk} f_\vk\right)$ ensures that configurations violating the constraint $f_o=0$ are assigned zero weight. The power spectrum is the 2-point correlator $\langle f f^*\rangle$ derived from the partition function $Z$ (by, e.g., adding a source term to the free energy). It reduces to the path integral \cref{eq:pathintegral} in the $V\to \infty$ limit.

The Dirac functional may be incorporated as a Fourier integral over ghosts $c$, $\bar{c}$ (which are both real numbers!).\footnote{As imposing the constraint is akin to spontaneously breaking the symmetry, the additional random numbers $c$ and $\bar{c}$ may be regarded, roughly speaking, as the accompanying ``Goldstone bosons''.} Note that $f(\mathbf{k})$ is a complex number, so the delta function must fix both its real and imaginary parts, provided the integration over $\vk$ is only over half of the volume (i.e. $\mathbb{R}^3$ modulo multiplication by $-1$). Thus
\begin{align}
  \langle f^*(\vp)f(\vq)\rangle &= \frac{1}{(2\pi)^2}\int\! \mathcal{D}f \int\!\mathrm{d}c\mathrm{d}\bar{c} \\ & \times \exp\left(\mathrm{i}\int_\vk (c\,\Re f(\vk) - \bar{c}\,\Im f(\vk))\right) \nonumber \\
  &\times f^*(\vp)f(\vq)\exp\left(-\int_\vk \frac{\abs{f(\vk)}^2}{2P_f(k)}\right) \nonumber 
\end{align}
where we have made explicit our particular choice of $f_o=0$ here.
Completing the square by defining
\begin{align}
    \label{eq:field_redefinitions_constrained}
        & f(\vk) = \tilde{f}(\vk) + \mathrm{i}P_f(k)\big(c-\mathrm{i}\bar{c}\big) \\ &
        f^*(\vk) = \tilde{f}^*(\vk) + \mathrm{i}P_f(k)\big(c+\mathrm{i}\bar{c}\big)
        \nonumber 
\end{align}
implies that the exponent becomes
\begin{equation}
    \label{eq:action_constrained_field}
  e^{-S[\tilde{f},\tilde{f}^*,c,\bar{c}]} = \exp\left(-\int_\vk \frac{\tilde{f}^*(\vk)\tilde{f(\vk)}}{2P_f(k)} + P_f(k)\frac{c^2+\bar{c}^2}{2}\right)\;.
\end{equation}
The functional Jacobian $\mathcal{D} f/\mathcal{D}\tilde{f}$ is, of course, unity. Thus, $\tilde{f}$ and $\tilde{f}^*$ are Gaussian random fields, like $f$ and $f^*$.

Terms linear in the ghosts vanish upon integration over $\mathrm{d}c\mathrm{d}\bar{c}$, whence
\begin{align}
    \label{eq:power_spectrum_constrained}
  \langle f^*(\vp)f(\vq)|0\rangle &\propto \int\!\mathcal{D}\tilde{f} \int\!\mathrm{d}c\mathrm{d}\bar{c}~e^{-S[\tilde{f},\tilde{f}^*,c,\bar{c}]} \\ & \times \left(\tilde{f}^*(\vp) \tilde{f}(\vq) - P_f(q)P_f(p)(c^2 + \bar{c}^2)\right) 
  \nonumber\;.
\end{align}
Both ghosts are Gaussian random variables with variance $2\left(\int_\vk P_f(k)\right)^{-1}$, where the factor of $2$ comes from integrating only over half of $\mathbb{R}^3$ in the exponent.
Therefore equation \cref{eq:power_spectrum_constrained} implies that
\begin{equation}
\label{eq:PSdeltaf0}
  \langle f^*(\vp)f(\vq)|0\rangle = (2\pi)^3\left[\delta^D\!(\vp-\vq)P_f(q) - \frac{P_f(q)P_f(p)}{(2\pi)^3\int_\vk P_f(k)}\right]
\end{equation} 
This equation has the usual conventions, where the integration $\mathrm{d}^3\vk$ is over all of $\mathbb{R}^3$.
Since the mean field identically vanishes when $f_o=0$, this last expression is precisely the (pseudo) power spectrum $P_{\delta f}(\vp,\vq)$ of the residual field $\delta f(\vx)$, see \cref{eq:PSdeltaf}.

The action in equation \cref{eq:action_constrained_field}, together with the definitions of the new fields, $\tilde{f}$, $\tilde{f}^*$, may serve to compute any constrained $n$-point correlation function in momentum space in the same manner. 
For example, taking the Fourier transform of equations \cref{eq:field_redefinitions_constrained} yields
\begin{equation}
    f(\mathbf{x}) = \tilde{f}(\mathbf{x}) + 2\mathrm{i}\xi_f(x) c.
\end{equation}
This expresses the real-space constrained field, $f$, in terms of the unconstrained Gaussian random field $\tilde{f}$, and an independent (ghost) random variable, $c$, which in normally distributed with variance $\sigma_c^2$. The only drawback of this definition is that $\tilde{f}$ is no longer a real field; but this is inconsequential.
    
\subsection{Including a survey mask}
\label{subsec:surveymask}

Ignoring the constant piece $\bar f$ in \cref{eq:Fsmall}, the Cartesian Fourier modes $F(\vk)$ are given by
\begin{equation}
    F(\vk) = \int\!\!d^3x\,{\cal W}(\vx)\big[\langle f\rangle(r)+\delta f(\vx)\big] e^{-\mathrm{i}\vk\cdot\vx}\;.
\end{equation}
Here again, the covariance $\langle F(\vk) F(\vk')\rangle$ of the Fourier modes $F(\vk)$ conditioned to $f({\bf 0})=f_o$ at the origin is the sum of the individual covariances arising from the mean field and from the residual. 

The first (disconnected) contribution is the product $\la {\cal W} f\ra(k)\la {\cal W} f\ra (k')$, which is
\begin{equation}
    \frac{f_o^2}{\sigma_f^4} \int_\vq P_f(q) {\cal W}(\vk-\vq) \int_{\vq'} P_f(q') {\cal W}(\vk'-\vq') \;, 
\end{equation}
where 
\begin{equation}
    {\cal W}(\vk)=\int\!\!d^3x\,{\cal W}(\vx)\, e^{-\mathrm{i}\vk\cdot\vx}
\end{equation}
is the Fourier transform of the survey window function ${\cal W}(\vx)$ (normalized to $\int\!\mathrm{d}^3x\,{\cal W}(\vx)=V$). In the particular case of a spherical mask function of radius $R$ centered at position $\vx_s$, such that the survey volume is $V=\frac{4\pi}{3}R^3$, ${\cal W}(\vk)$ is given by
\begin{equation}
    {\cal W}(\vk) = 3V \frac{j_1(kR)}{(kR)} e^{-\mathrm{i}\vk\cdot\vx_s}\;.
\end{equation}
The second (connected) contribution,$\la({\cal W} \delta f)(\vk)({\cal W}\delta f)(\vk')\ra$, is given by
\begin{equation}
    \int_\vq\int_{\vq'}\,P_{\delta f}(\vq,\vq') \,{\cal W}(\vk-\vq){\cal W}(\vk'-\vq') \;.
\end{equation}
Using \cref{eq:PSdeltaf}, this becomes
\begin{multline}
    \int_\vq P_f(q) {\cal W}(\vk-\vq){\cal W}(\vk'+\vq) \\ 
    - \frac{1}{\sigma_f^2}\int_\vq P_f(q) {\cal W}(\vk-\vq)\int_{\vq'}P_f(q'){\cal W}(\vk'-\vq') \;.
\end{multline}
Adding up these two contributions and specializing to the case $\vk=-\vk'$, the covariance of $F(\vk)F(-\vk)=F(\vk)F^*(\vk)$  reads
\begin{align}
\big\langle\big\lvert F(\vk)\big\lvert^2\big\rangle
    &= \int_\vq P_f(q) \big\lvert{\cal W}(\vk-\vq)\big\lvert^2 + \frac{1}{\sigma_f^2}\left(\frac{f_o^2}{\sigma_f^2}-1\right) \nonumber \\
    &\quad \times \int_\vq P_f(q){\cal W}(\vk-\vq)\int_{\vq'}P_f(q'){\cal W}(-\vk-\vq') 
    \nonumber \;,
\end{align}
which leads immediately to \cref{eq:Pestimated}.
Note that this expression depends solely on $k=|\vk|$ if the survey window function is isotropic.

In the limit $V\to\infty$, ${\cal W}(\vk)$ is narrowly peaked around $\vk=0$. Consequently, both  ${\cal W}(\vk)$ and $V^{-1}|{\cal W}(\vq-\vk)|^2$ asymptote to $(2\pi)^3 \delta^D(\vq-\vk)$. While this is obvious for the former, note that
\begin{align*}
    V^{-1}|{\cal W}(\vk)|^2 &= V^{-1}\int\!\!d^3x\int\!\!d^3y\, {\cal W}(\vx){\cal W}(\vy) e^{\mathrm{i}\vk\cdot(\vx-\vy)} \\
    &\approx V^{-1}\int_V\!\!d^3x\int_V\!\!d^3r\,e^{\mathrm{i}\vk\cdot\vr} \nonumber \\
    &= \int_V\!\!d^3r\,e^{\mathrm{i}\vk\cdot\vr} \nonumber \\
    &\stackrel{V\to\infty}{=} (2\pi)^3 \delta^D\!(\vk) \;.
\end{align*}
This demonstrates that $\hat P_F(\vk)$ converges towards \cref{eq:Plargevolume} for large survey volumes. 

The variance var$(\hat P)_F$ can be computed from the covariance
\begin{align}
    {\rm Cov}\big[\hat P(\vk),\hat P(\vk')\big] &= \big\langle F(\vk)F(-\vk')\big\rangle
    \big\langle F(-\vk)F(\vk')\big\rangle \\
    &\quad +\big\langle F(\vk)F(\vk')\big\rangle\big\langle F(-\vk)F(-\vk')\big\rangle 
    \nonumber \\
    &\quad - 2 \big\langle F(\vk)\big\rangle \big\langle F(-\vk)\big\rangle \big\langle F(\vk')\big\rangle \big\langle F(-\vk')\big\rangle \nonumber \;.
\end{align}
Substitution of the previous results into this relation leads to \cref{eq:varPestimated}.

\subsection{Quantum-Mechanical perspective}
\label{appendix:quantum}

From a quantum mechanical point of view, the statistics of the random field $f$ stem from the statistics of a certain quantum state $\rho$, that describes the state of $f$, \emph{qua} a quantum field. In this appendix we show how the constrained statistics described above arise from a quantum-mechanical point of view. Let us construct Hilbert spaces on $t = \textrm{const}$ hyper-surfaces (equivalently $\eta = \textrm{const}$ hyper-surfaces), and let $\ket{\psi(t_0)}$ be the wave-function of the Universe at $t=t_0$, where $t_0$ is the time at which the observer $O$ makes the measurement. This decomposition is gauge dependent, but the existence of the observer at $\vx_0$ is coordinate independent of course. It is clear that not all initial conditions give rise to an observer $O$ situated at $\vx_0 = \boldsymbol{0}$, so one can decompose $\ket{\psi(t_0)}$ into
\begin{align}
    \ket{\psi(t_0)} &= \int\! \mathrm{d}^3\vx \; \Big[a(\vx)\ket{0_{\vx}}\ket{\phi_0(t_0,\vx)} \nonumber \\ & \qquad + b(\vx)\ket{1_{\vx}}\ket{\phi_1(t_0,\vx)}\Big] \;,
\end{align}
where $a(\vx)$ and $b(\vx)$ are complex-valued functions; and $\ket{1_{\vx}}$, $\ket{0_{\vx}}$ formally correspond to there being an observer at $\vx_0$ at $t_0$, or not. When $O$ performs a measurement, one first has to project $\ket{\psi(t_0)}$ on the sub-space corresponding to the existence of $O$ at $\vx_0$. That is, \emph{before} $O$ measures any operator, an additional projection is made, of whether $O$ exists or not (cf. \citealt[\S II.C.3]{Tegmark:2011pi}). This amounts to acting with $\ket{1_{\vx_0}}\bra{1_{\vx_0}}$ and afterwards normalizing the result. This is followed by tracing out degrees of freedom outside the horizon (as they cause decoherence of the state inside it); denoting this partial trace by $\Tr_{\textrm{oh}}$, one finds 
\begin{equation}
    \rho(t_0,\vx_0) \propto \Tr_{\textrm{oh}}\left(\ket{1_{\vx_0}}\bra{1_{\vx_0}}\ket{\psi(t_0)}\bra{\psi(t_0)}\ket{1_{\vx_0}}\bra{1_{\vx_0}}\right) \;,
\end{equation}
where the density matrix $\rho$ describes the probability to have a certain outcome conditioned to the existence of $O$. Its normalization is fixed by requiring that its trace be $\Tr\;\rho(t_0,\vx_0) = 1$.

For simplicity take $f$ to be the galaxy over-density $\delta_g$, which is ultimately the observed quantity (it is related to the dark matter fields via equation \eqref{eq:GRng} as explained above in Section \ref{subsec: Galaxy number counts}). It is clear that its value at $\vx_0$ affects the possibility of the existence of an observer, and as it is locally measurable, it must be fixed prior to a cosmological observation to $f_o$ (cf. \citealt{reischke/schaefer/etal:2019}). 
When an observation is conducted, it is automatically correlated with past or present other observations, e.g. the very existence of the observer, and the value of $\delta_g$ at its location. Not conditioning on them would amount to a mis-representation of the probability distribution of the current observation's outcomes.
This implies a further projection (and subsequent normalization) acting on $\rho(t_0,\vx_0)$, that projects on the sub-space that has $f(\vx_0) = f_o$; let $\rho(t_0,\vx_0,f_o)$ be the result. 
If $O$ now proceeds to measure any Hermitian operator $A$ corresponding to a cosmological observable (inside the horizon), its expectation value (i.e. its ensemble average) is given by $\langle A \rangle = \Tr[A\rho(t_0,\vx_0,f_o)]$. This is not the same as an average over all initial field configurations -- the latter would correspond to $\bra{\psi(t_0)}A\ket{\psi(t_0)}$. The former expectation value is precisely the one obtained by using the constrained ensemble adopted in this paper.

\subsection{Local primordial non-Gaussianity}

For simplicity, we consider the initial spatial curvature $\phi_i(\vx)$ rather than the present epoch $\phi(\vx,\eta)$.
In the presence of a local primordial non-Gaussianity, $\phi_i(\vx)$ can be expressed as 
\citep{salopek/bond:1990,gangui/etal:1994,komatsu/spergel:2001}
\begin{equation}
    \phi_i(\vx) = \phi_G(\vx) + f_\textrm{NL}\big(\phi_G^2(\vx) - \langle\phi_G^2\rangle\big),
\end{equation}
where, as before, the expectation value is a constrained one. The Gaussian piece $\phi_G(\vx)$ has a power-spectrum $P_G(k)\propto k^{n_s-4}$ and a variance $\sigma^2_G$ computed from all the modes within the observable Universe. A constraint on $\phi_i(\vx)$, $\phi_i({\bf 0}) = \phi_{oi}$, straightforwardly translates into a constraint on $\phi_G(\vx)$, \emph{viz.}, $\phi_G({\bf 0}) = \phi_{oG}$. We calculate the resulting non-Gaussian correction to the constrained power spectrum $\langle \phi_i(\vk)\phi_i(\vk')\rangle$ using the path integral approach. This involves 2-, 3- and 4-point functions of $\phi_G$, each of which are calculated using equations \eqref{eq:field_redefinitions_constrained}, for $\phi_G$, by defining $\tilde{\phi}_G$ in the same manner. Then, $\tilde{\phi}_G(\vk)$ is an un-constrained Gaussian random field, with power-spectrum $P_G(k)$. This implies that 
\begin{align}
    \langle \phi_i(\vk) &\phi_i^*(\vk')\rangle = \langle \phi_G(\vk)\phi^*_G(\vk')\rangle \\ & 
    + f_{\rm NL} \int_\vq \langle \phi_G(\vq)\phi_G(\vk - \vq)\phi_G^*(\vk')\rangle \nonumber \\ & 
    + f_{\rm NL} \int_{\vq} \langle \phi_G(\vk)\phi_G^*(\vq)\phi_G^*(\vk'-\vq)\rangle \nonumber \\ &
    + f_{\rm NL}^2 \iint_{\vq_1,\vq_2} \langle \phi_G(\vq_1)\phi_G(\vk - \vq_1)\phi_G^*(\vq_2)\phi_G^*(\vk'-\vq_2)\rangle \nonumber \;.
\end{align}
Evaluating these correlators in the same manner as above yields, after some algebra,
\begin{align}
    \label{eq:covphii}
    \langle \phi_i(\vk)\phi_i^*(\vk')\rangle & = (2\pi)^3\delta^D\!(\vk-\vk')P_G(k) \\ & \quad 
    + \frac{1}{\sigma_G^2}\left(\frac{\phi_{oG}^2}{\sigma_G^2}-1\right) P_G(k)P_G(k') \nonumber  
    \\ & \quad \bigg. + f_\textrm{NL}^2(2\pi)^3\delta^D\!(\vk-\vk')\,\Big(P_2(\vk)+P_2(\vk')\Big) \nonumber \\ & \quad 
    +\frac{4 f_\textrm{NL}\phi_{oG}}{\sigma_G^2}P_G(k)P_G(k') \nonumber  \\ & \quad
    - \frac{2f_\textrm{NL}\phi_{oG}}{\sigma_G^4}\Big(P_G(k)P_2(\vk)+P_G(k')P_2(\vk')\Big) \nonumber  \\ & \quad
    + \frac{4f_\textrm{NL}^2\phi_{oG}^2}{\sigma_G^2}\left(\frac{P_3(\vk,\vk')}{\sigma_G^2} - \frac{P_2(\vk)P_2(\vk')}{\sigma_G^4}\right) \nonumber \\ & \quad
    -\frac{4f_\textrm{NL}^2}{\sigma_G^2}P_3(\vk,\vk') + \frac{2f_\textrm{NL}^2}{\sigma_G^4}P_4(\vk,\vk') \nonumber \;,
\end{align}
where 
\begin{align}
    & P_2(\vk) = \int_\vq P_G(q)P_G(|\vk-\vq|) \\ &
    P_3(\vk',\vk) = \int_\vq P_G(q)P_G(|\vk-\vq|)P_G(|\vk'-\vq|) \nonumber \\ &
    P_4(\vk',\vk) = \int_{\vq_1}\int_{\vq_2} P_G(q_1)P_G(q_2)P_G(|\vk-\vq_1|)P_G(|\vk'-\vq_2|) 
    \nonumber \;.
\end{align}
The third term in the right-hand side of \cref{eq:covphii} is the unconstrained, non-Gaussian contribution to the covariance of $\phi_i(\vk)$, which is of order $\sim (f_\text{NL}\sigma_G)^2$ relative to the first term. The fourth term is the leading, constrained contribution arising from the local primordial non-Gaussianity. 
For a typical fluctuation $\sigma_{oG}\sim \sigma_G$, the amplitude of this correction relative to the constrained, Gaussian piece (the second term) is of order $\sim f_\text{NL}\sigma_G$. For $f_\text{NL}\sim 1$ and $\sigma_G\lesssim 10^{-4}$, both contributions can be ignored and, thereby, all the remaining terms proportional to $f_\text{NL}$ and $f_\text{NL}^2$ as well. A proper calculation taking into account the transfer function would not alter our conclusion.

\end{document}